\documentclass[twocolumn]{aastex63}
\usepackage{natbib}
\usepackage{amsmath} % or simply amstext

\newcommand{\msun}{$M_{\odot}$}

\newcommand{\mstar}{$M_*$}
\newcommand{\mbh}{$M_{\rm BH}$}
\newcommand{\jwst}{\ensuremath{JWST}}

\begin{document}

\title{An empirical approach to selecting the first growing black hole seeds with JWST/NIRCam}
\author[0000-0003-4700-663X]{Andy D. Goulding}
\affiliation{Department of Astrophysical Sciences, Princeton University,Princeton, NJ 08544, USA}
\author[0000-0002-5612-3427]{Jenny E. Greene}
\affiliation{Department of Astrophysical Sciences, Princeton University,Princeton, NJ 08544, USA}

%\date{Aug 2022}
\submitjournal{ApJ}

\begin{abstract}
The James Webb Space Telescope (JWST) will have the sensitivity to detect early low-mass black holes as they transition from ``seeds'' to supermassive black holes (BHs). Based on the JAGUAR mock catalog of galaxies, we present a clean color selection that takes advantage of the unique UV slope of accreting supermassive black holes with relatively low mass and high accretion rates. We show that those galaxies hosting $\sim 10^{6} M_{\odot}$ BHs radiating at $>10\%$ of their Eddington luminosity separate in color space from inactive systems for a range of host stellar masses. Here we propose a set of 3-band, 2-color selection boxes (with 90\% completeness; 90\% purity; balanced purity/completeness) with JWST/NIRCam to identify the most promising growing BH candidates at $z \sim 7-10$. 
\end{abstract}

\keywords{Active galactic nuclei (16), High-redshift galaxies (734), Intermediate-mass black holes (816), Early universe (435)}

\section{Introduction}

For years, we have mused about the origin story of supermassive black holes \citep[BHs; see review by ][]{Inayoshi:2020}. The boundary conditions are clear: we must grow the first billion solar mass BHs only a couple hundred million years after the Big Bang \citep[e.g.,][]{banadosetal2018}. BH seeds may form through the death of the first generation (Population III) stars \citep[e.g.,][]{loebrasio1994,brommloeb2003}, but then some must grow at super-Eddington rates to produce the first quasars. Alternatively, seeds may form through dynamical processes in the centers of dense stellar clusters \citep[e.g.,][]{portegieszwartetal2002}, either rapidly through the formation of a super-massive star \citep[e.g.,][]{freitagetal2006,devecchivolonteri2009}, or more slowly through merging of compact objects \citep[e.g.,][]{milleretal2002}. In principle such mergers can make $\sim 1000$~\msun\ seeds, provided that the merger products are not ejected first \citep[e.g.,][]{holley-bockelmannetal2008}. 

Dense and massive star clusters may also provide a fruitful environment for light (10-100~\msun) seeds to grow, through some combination of accretion, dynamical interactions, and tidal disruption \citep[e.g.,][]{stoneetal2017,Kroupa2020,Natarajan:2021}. Even more massive seeds may collapse directly from low angular-momentum gas without forming stars first; so-called ``direct-collapse'' models in principle form $10^4-10^6$~\msun\ seeds \citep[e.g.,][]{koushiappasetal2004,lodatonatarajan2006,begelman2010}, but debate persists about how rare the perfect conditions for collapse may be \citep[e.g.,][]{habouzitetal2016}. See the comprehensive reviews in \citet{Inayoshi:2020} and \citet{Volonteri:2021}.

Up until now, it has not been possible to detect the emission from seed BHs directly. { Thus, observational constraints have focused on two areas: detecting the earliest accreting black holes on the one hand, and searches for relic seed BHs locally on the other. Both paths are important. We are able to detect the most luminous and massive accreting BHs at $z>6$ \citep[e.g.,][]{Mortlock:2011,Banados:2018,Fan:2019,Wang:2021,Harikane:2022}, but their number densities are very low compared to the full supermassive BH population. Thus, while the existence of these quasars at early times does provide one boundary condition on seed formation, it is not necessarily the case that the process forming these rare beasts is the universal seeding mechanism. 

As pointed out by \citet{volonterietal2008} and others, relic seed BHs also contain clues as to the nature of seed formation.} BHs with masses of $\sim 10^5$~\msun\ may be found dynamically within $D \approx 3-5$~Mpc \citep[e.g.,][]{gebhardtetal2002,sethetal2014,nguyenetal2018,pechettietal2017}, or via the detection of signatures of active galactic nuclei (AGN) from optical spectroscopic signatures \citep[AGN; e.g.,][]{filippenkosargent1989,reinesetal2013}, optical variability \citep[e.g.,][]{baldassareetal2016,Burke:2022}, X-ray \citep[e.g.,][]{pardoetal2016,mezcua2019}, or mid-infrared spectroscopic signatures \citep[][]{satyapaletal2007,gouldingetal2010}. For a comprehensive review, see \citet{Greene:2020}.

Everything has changed with the successful launch of the \emph{James Webb Space Telescope} (\jwst). This infrared-optimized space telescope is sensitive enough to detect Eddington-limited $10^6$~\msun\ BHs to $z \approx 10$. Therefore, we urgently need an understanding of how to observationally identify these candidates within the \jwst\ data. A number of theoretical papers have explored various seed formation scenarios along with predictions for observable signatures with \jwst\ in continuum-selection \citep[e.g.,][]{Volonteri:2017,natarajanetal2017,pacuccietal2018,Valiante:2018,Latif:2018,Whalen:2020,Yung:2021} as well as emission-line selection \citep[e.g.,][]{Nakajima:2022}. 

We are motivated to revisit the observability of accreting seed BHs with \jwst\ for two reasons. First of all, nearly all of the existing predictions are based on some combination of \jwst/MIRI, deep X-ray data, and/or mid-to-far infrared luminosity. These latter two are out of reach for the foreseeable future (as we will argue below) while MIRI imaging is unlikely to be as deep as NIRCam imaging at first. Thus, we set out to create NIRCam-specific search criteria to be deployed in the coming months as deep field data at the requisite depths become available. Secondly, we employ a complementary empirical approach to the modeling papers referenced above.

We combine well-motivated mock galaxy catalogs built for \jwst\ \citep{Williams:2018} with observed lower-redshift AGN templates for low-mass, high Eddington-ratio galaxies to make detailed predictions for \jwst/NIRCAM color selection. We assume a sensitivity of AB$=30$~mag, motivated by upcoming deep field surveys like the Ultra-deep NIRCam and NIRSpec Observations Before the Epoch of Reionization (UNCOVER; PIs Labbe and Bezanson). We focus on the minimum detectable BH, with \mbh$\approx 10^6$~\msun\ radiating at its Eddington limit, at $z \sim 7-10$.

We will describe our assumed spectral energy distribution (SED) for accreting BHs in \S \ref{sec:AGNSED}, and the mock galaxy catalog in \S \ref{sec:galaxy}. Putting these together, we demonstrate the color space occupied by low-mass AGN in \S \ref{sec:colorbox}, and then discuss the possible implications of finding these populations in \S \ref{sec:discussion}. Throughout, we assume a standard flat $\Lambda$CDM cosmology with $H_0 = 67.7$~km~s$^{-1}$~Mpc$^{-1}$ and $\Omega_M = 0.31$ \citep{Planck:2020}.

\section{The AGN Spectral Energy Distribution}
\label{sec:AGNSED}

%%%%%%%%%%%%%%%%%%%%%%%%%%%%%%%%%%
\begin{figure*}
\includegraphics[width=0.98\textwidth]{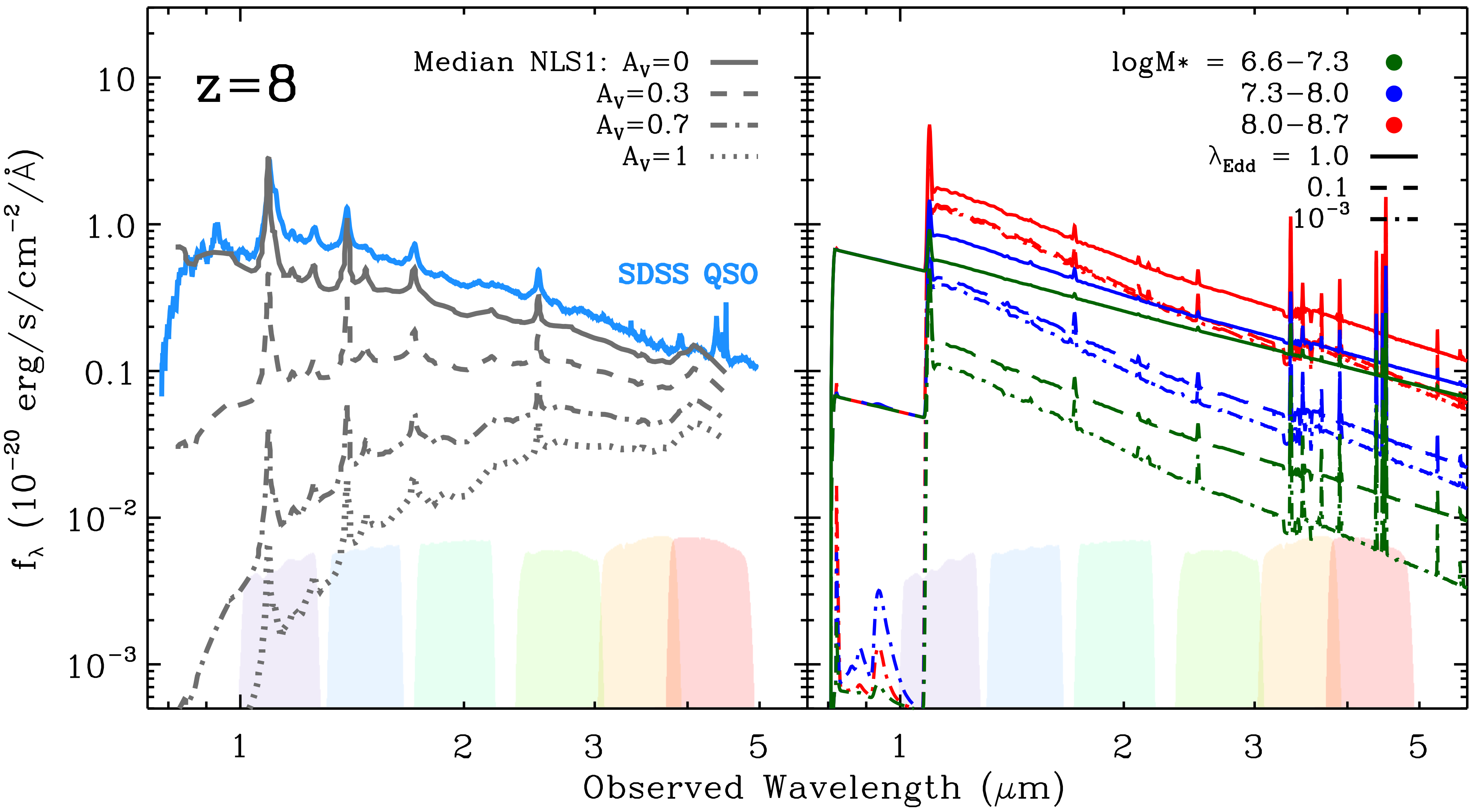}
\caption{Examples of the assumed spectral energy distributions in the NIRCAM bands. Left panel provides the ``Median NLS1'' spectrum from \citet{Constantin:2003}, which has $f_{\lambda} \propto \lambda^{-1.21}$, and is flatter and more shallow than the median quasar spectrum of substantially more massive BHs typically observed in the SDSS (blue; Vanden Berk et al. 2001). The effect of increasing dust extinction ($A_V = 0.3, 0.7, 1.0$) is additionally shown. Right panel shows the galaxy SEDs extracted from the JAGUAR mock combined with the NLS1-inspired AGN model for a range of galaxy stellar masses (log$M_* = 6.6-7.3, 7.3-8.0, 8.0-8.7$) along with a range of Eddington ratios for the accreting BHs ($\lambda_{\rm Edd} = 10^{-3}, 0.1, 1.0$). For an assumed \mbh\ of $5 \times 10^6$~\msun, we expect to detect AGN with UV luminosities in the range $M_{\rm UV,1450} \sim -16.6$ to $-19.1$.}
\label{fig:sed}
\end{figure*}
%%%%%%%%%%%%%%%%%%%%%%%%%%%%%%%%%%

The first essential ingredient for our predictions is a template accreting low-mass BH. We only have the sensitivity to detect BHs at the upper end of our mass range of interest (\mbh$=10^4-10^6$~\msun) and only those at (or above) their Eddington limit. In this work, we will assume no BHs exceed this limit, but we note the growing theoretical literature suggesting that the Eddington luminosity may not be a limit \citep[e.g.,][]{jiangetal2019}. Future work may consider how SED changes at super-Eddington rates would translate into different observability constraints.

Under the basic assumptions of the $\alpha-$disk model \citep{Shakura:1973}, the accretion disk emits as a multi-temperature black-body and the peak temperature scales as \mbh$^{-1/4}$. Therefore, lower-mass BHs peak at hotter temperatures or higher frequencies. Well-used SED fitting codes such as {\tt CIGALE} tend to use AGN templates (such as those from the Skirtor libraries; e.g., \citealt{Stalevski:2016}) with fixed piecewise slopes that are motivated by average quasar templates{, such as those derived from the Sloan Digital Sky Survey \citep{VandenBerk:2001}}. By design, these AGN templates have relatively steep UV spectra that are more typical of high-mass BHs, and may not be well suited to the lower mass population that is expected in the early Universe. Lower-mass BHs will have disks that peak at higher energies. \citet{Volonteri:2017} adopt theoretical disk models that account for lower-mass and thus hotter disks. { However, models can struggle to match observed quasars in detail \citep[e.g.,][]{Davis:2007}, and so we prefer to use empirical templates comprised of rapidly accreting low-mass BHs (albeit at lower redshift).}

The best local analogs for rapidly growing, low-mass BHs are the narrow-line Seyfert 1 galaxies \citep[NLS1; e.g., ][]{Osterbrock:1985}. NLS1s are defined by their relatively narrow broad lines, which is thought to indicate low \mbh, while the corresponding relatively high luminosities indicate objects that are approaching their Eddington limits \citep[see review in ][]{Komossa:2008}. Of course, these BHs are not perfect analogs for early $\sim 10^6$~\msun\ BHs (e.g., the metallicity of the surrounding gas in high-redshift AGN will likely be lower, while the gas densities may be higher). However, to zeroth order, the shape of the UV continuum should be predominantly determined by \mbh, and so UV observations of NLS1s likely provide the best available template for the UV emission of early growing BHs.

\citet{Constantin:2003} present a composite UV spectrum of NLS1s at low redshift. They fit a continuum slope between 1100 and 4000\AA\ of $f_{\nu} \propto \nu^{-0.79}$, as shown in Figure \ref{fig:sed}. We adopt this power-law model with the slope measured by \citet{Constantin:2003}. { When normalized at rest frame $\lambda \sim 5000$\AA, we show that unlike the median SDSS quasars, our adopted AGN SED does not turn over in the far-UV, owing to the lower \mbh\ of NLS1s, and the shallower slope produces a factor $\sim$2 lower flux at 2500\AA\ when compared to the SDSS quasars.}

The environments hosting early BH growth may also be dust rich \citep{natarajanetal2017}, and hence we investigate the effect of dust extinction on our adopted AGN model. Given the rest-frame wavelengths, even mild attenuation ($A_{\rm V} < 1$) towards the AGN will have a significant effect on the observed flux. Here we adopt a typical SMC extinction curve (e.g., \citealt{Hopkins:2004}), and investigate a range of $A_{\rm V}$ from 0.3-1.0.

%%%%%%%%%%%%%%%%%%%%%%%%%%%%%%%%%%
\begin{figure*}[t]
\centering
\includegraphics[width=0.85\textwidth]{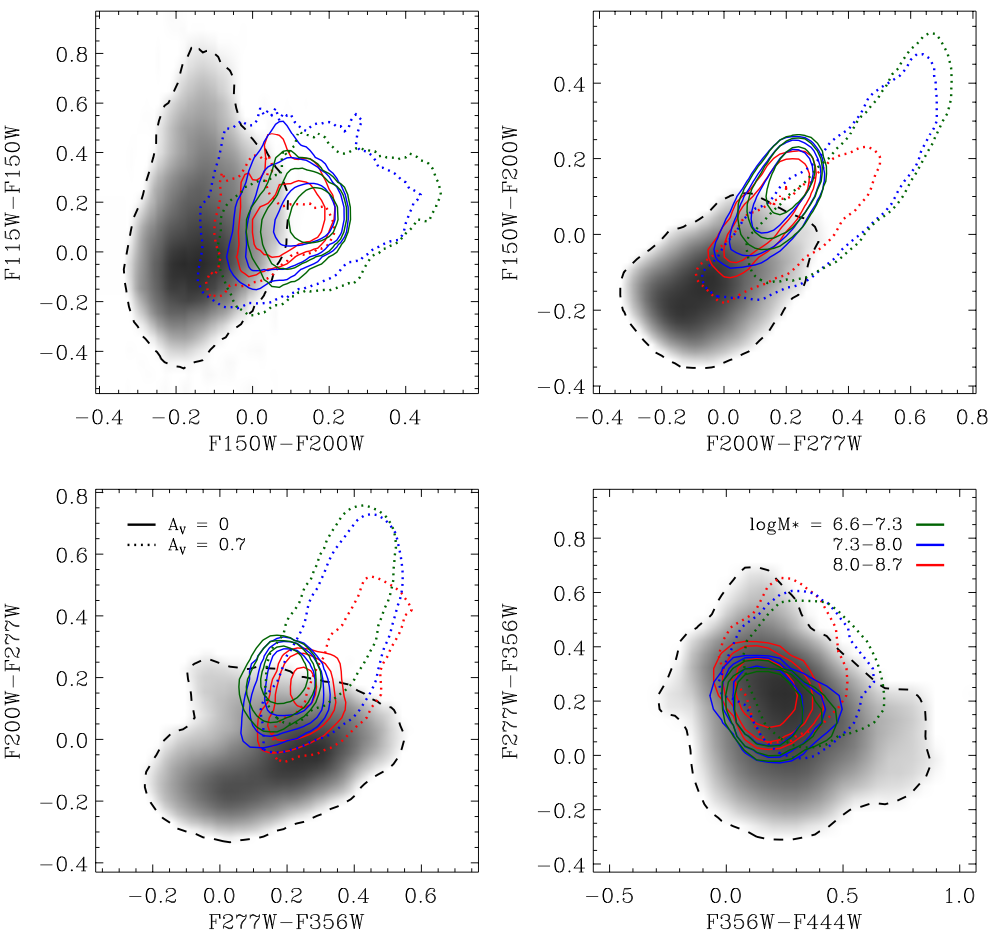}
\caption{Comparison between the color distribution of the mock galaxies (greyscale) and galaxy+AGN models (colored contours) for \mbh$=5 \times 10^6$~\msun\ with a log-normal distribution in Eddington ratio between $0.1-1$. Solid contours provide the 67,90 and 99\% distributions for AGN with no applied dust extinction ($A_{\rm V}=0$) and dotted contours provide the 99\% distribution for AGN with $A_{\rm V}=0.7$. Galaxy masses run from \mstar$\sim 10^{6.5}-10^9$~\msun, and 25\% of galaxies harbor BHs accreting at or above 10\% of their Eddington limit.
\label{fig:colors}}
\end{figure*}
%%%%%%%%%%%%%%%%%%%%%%%%%%%%%%%%%%

\section{Mock Galaxy Catalog}
\label{sec:galaxy}

The other critical ingredient for our predictions is to have realistic galaxy colors. There is some chance that some seed BHs form and grow dramatically before the galaxy around them \citep[e.g.,][]{natarajanetal2017}. That would be the easiest case for detecting such systems, although we note that the BH would have to reach above \mbh$=10^6$~\msun\ to be detectable in typical \jwst\ deep fields (without gravitational lensing). Otherwise, the galaxy will contribute a comparable amount of light in the UV as the BH. Thus, the question becomes, if we add this empirically motivated power-law continuum to expected galaxy model spectra, is there a tell-tale signature in the colors for the presence of an accreting BH.

We draw galaxy models from the \jwst\ mock catalog the JAdes extraGalactic Ultradeep Artificial Realizations or JAGUAR \citep{Williams:2018}. This mock catalog was built with the upcoming program the \jwst\ Advanced Deep Extragalactic Survey (JADES) in mind. We will briefly describe salient aspects of the model here, but for details see Williams et al.\ (2018).

At $z > 4$ the JAGUAR number counts are based on the measured UV luminosity function \citep{Bouwens:2015,Oesch:2018} and an empirically motivated relationship between UV luminosity and stellar mass, meant to encapsulate the variety of star formation histories and dust contents that determine the UV mass-to-light ratio in practice. For $z<4$, the 3D-HST \citep{Brammer:2012} catalog is used to characterize the $M_{\rm UV}-M_*$ relation. A constant slope is assumed, and the normalization is allowed to vary with redshift. At $z>4$, { it is no longer possible to independently constrain the slope and zeropoint of the relation due to limited dynamic range in galaxy luminosities and the challenges of removing emission line contamination to measure reliable stellar mass \citep[e.g.,][]{Labbe:2013,Stark:2013}. Therefore, a constant slope is assumed at higher redshift, while observed $z>4$ galaxies at the bright end $M_{\rm UV} = -20$~mag set the normalization \citep{Stark:2013}}. For similar reasons, the scatter in UV luminosity at fixed mass is measured at $z<3$ and fixed at higher redshift. 

Finally, a theoretical floor in the mass-to-light ratio is derived from the stellar population synthesis code BEAGLE \citep{Chevallard:2016} and no galaxies are allowed to exceed that UV luminosity threshold. The spectral slope of each galaxy in the UV is drawn from an empirical relation between the slope $\beta$ and $M_{\rm UV}$ \citep{Bouwens:2009,Bouwens:2014}, again with a theoretical limit on the bluest possible O-stars imposed. The relation is not evolved for $z>8$ galaxies. A constant scatter of $\sigma_{\beta} = 0.35$ is used across all redshifts \citep[e.g.,][]{Bouwens:2012,Mehta:2017}, with a limiting slope of $\beta = -2.6$. Finally, star-formation history parameters corresponding to these UV colors are assigned from a library of mock galaxy star-formation histories with imposed relationships between stellar mass, star-formation history, gas-phase metallicity, and dust attenuation. For our purpose, the detailed star-formation histories matter less than the range of $\beta$ at low mass, since it is the differences between $\beta$ and the AGN UV slope $\alpha$ defined in \S \ref{sec:AGNSED} that determine detectability. 

Williams et al.\ extensively validate that their galaxies are consistent with measured UV luminosity functions, star-formation rate density measurements, and emission-line equivalent widths measured from \emph{Spitzer} (their \S 6). As emphasized by Williams et al., the $M_{\rm UV}-M_*$ relation on which the JAGUAR galaxy population is built is an extrapolation, which therefore bakes in a number of selection effects in current samples. Our approach using the JAGUAR model that we adopt here is very similar in spirit to the galaxy model adopted by \citet{Volonteri:2017}. Others of the prediction papers extract star-formation histories from their models and then model the spectral energy distribution using synthesis models and photoionization modeling \citep[e.g.,][]{natarajanetal2017,Valiante:2018,Barrow:2018}.

New constraints on the mass functions of galaxies with $z>4$ are coming soon directly from \jwst\ data, and our methods can easily be extended when the true color space is empirically determined.

\section{Identifying AGN}
\label{sec:AGNid}

We select a sample of potential host galaxies by imposing three magnitude cuts, assuming photometry with a $5 \, \sigma$ limit of 30 AB mag. We enforce F070W, F090W$<1.4$~nJy, corresponding to $z>6.6$. Then we select only those galaxies with F200W$>3.7$~nJy to ensure a significant detection at redder wavelengths. These cuts together yield mock galaxies with median $\langle$\mstar$\rangle= 2 \times 10^7$~\msun, and nearly all lie within $6.6 < z < 10$ (0.5\% are at lower redshift). This redshift range ensures that we will have adequate spectral coverage in NIRCam imaging alone for low-mass AGN selection, and is motivated by upcoming \jwst\ deep fields \citep[e.g.,][]{Williams:2018}. 

As mentioned above, we model BHs with \mbh~$\sim 10^6-10^7$~\msun, as these are detectable with limiting $m_{\rm AB} \approx 30$~mag flux limits that characterize many upcoming \jwst\ deep fields. For each model BH, we draw a luminosity from a log-normal distribution, peaking near the Eddington limit and extending to $\sim L_{\rm bol}/L_{\rm Edd} \sim 0.1$. For lower Eddington ratios, we both do not have a suitable SED, nor are they luminous enough to be detected above the host galaxy dilution, even for an $A_{\rm V}=0$ scenario. { The exact parameterization of the Eddington ratio distribution has little impact on our results since we predict that, with the exception of massive BHs potentially hosted in extremely low mass galaxies ($< 10^6$\msun), only BHs with $L_{\rm bol}/L_{\rm Edd} > 0.1$ will be detectable based upon the planned flux limits of upcoming JWST survey fields.} We model the SED as a simple power-law, with the slope taken from Constantin et al. We do not include emission lines, but we can take the equivalent width (EW) of the broad CIV line in the AGN within the low-mass AGN NGC 4395 and estimate how much color change we might expect from emission lines. \citet{petersonetal2005} find a \ion{C}{4} equivalent width of $\sim 65$\AA, which would roughly boost the F150W band by $\sim 20-30\%$. Since this addition would only serve to move the AGN locus further from the galaxy locus in our selection (\S \ref{sec:colorbox}), we do not include it. No other significant broad lines fall within the main bands required for our selection (F150W, F200W, F277W, F356W, F444W).

%%%%%%%%%%%%%%%%%%%%%%%%%%%%%%%%%%
\begin{figure*}
\includegraphics[width=0.95\textwidth]{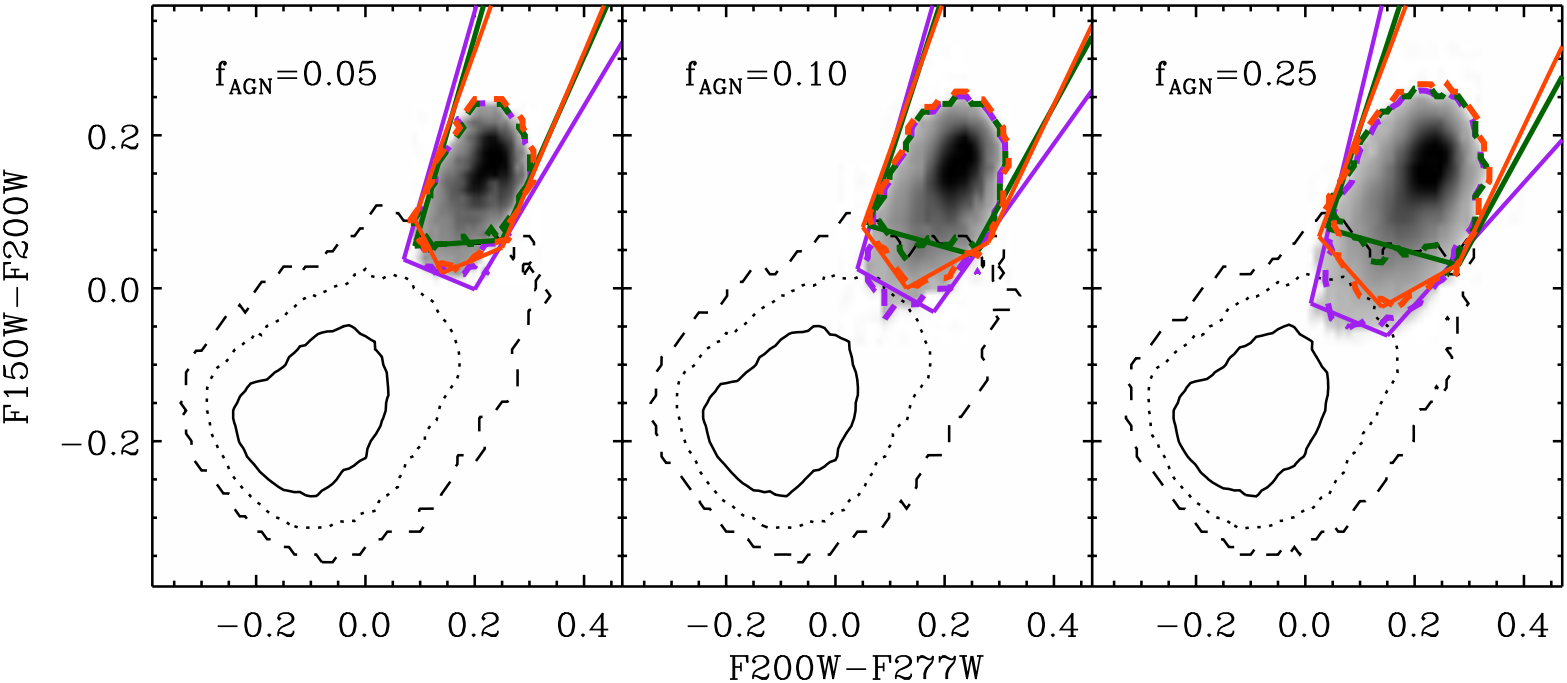}
\caption{F150W-F200W vs F200W-F277W color-color diagram and proposed AGN selection boxes assuming AGN fractions of $f_{\rm AGN} = {0.05,0.1,0.25}$ for three potential science drivers: 90\% purity (orange); 90\% completeness (purple); balanced 75\%--75\% completeness to purity ratio (green). Greyscale region represents the distribution of AGN in our mock samples, and black contours provide the distribution of galaxy colors in the JAGUAR simulation with F090W,F115W$<$1.4 nJy and F200W$>$3.7~nJy.}
\label{fig:colorbox}
\end{figure*}
%%%%%%%%%%%%%%%%%%%%%%%%%%%%%%%%%%

Finally, to calculate the combined AGN+galaxy colors, we draw from the JAGUAR hosts, add in the AGN model, and perturb the observed magnitudes assuming a 30 mag $5 \, \sigma$ detection. This set of simulations leads to the galaxy regions shown in grey-scale in Figure \ref{fig:colors}, while the AGN+galaxy models are shown in green, blue, red for log~\mstar$=6.6-7.3,7.3-8.0, 8.0-8.7$~\msun. We clearly see that the AGN+galaxy slope is redder than the galaxy slopes, causing the AGN to peak up in color space as a red object in F150W-F200W, and F200W-F277W. This difference in color distributions will allow us to select low-mass BH candidates at $z> 7$. We further investigate the effect of dust attenuation towards the central BH on the observed NIRCam colors. As stressed above, even mild attenuation at rest-frame UV wavelengths has a dramatic effect on the AGN continuum. Here we specifically assume that the attenuation is in the vicinity of the BH, and does not affect the host galaxy colors. We show in Fig.~\ref{fig:sed} that an $A_{\rm V} \sim 0.7$ reduces the flux of the AGN continuum by an order of magnitude in the F200W band at $z\sim8$, which for even the Eddington-limited case, produces a comparable flux to the lowest mass galaxies found in JAGUAR at these redshifts. Hence, host galaxy dilution becomes a limiting issue for the identification of AGN signatures in the NIRCam bands. However, due to the predominately blue colors of the host galaxies, we show that the redder colors of dust attenuated AGN extend the parameter space to redder NIRCam colors in almost all of the bands, although we caution that this is at the expense of signal to noise owing to the lower observed fluxes. By contrast, we show that for the longer wavelength colors available with NIRCam, the galaxies hosting AGN do not appreciably differ from the pure galaxy population in the unobscured $A_{\rm V}=0$ scenario, and hence, may not be readily useful in selecting those galaxies host growing BH seeds. We also note that at these early epochs, a non-negligible fraction of the growing BH population may also be heavily obscured, potentially even Compton-thick. While such a population will be missed at rest-frame UV/optical wavelengths due to heavy dust obscuration ($A_{\rm V} \gg 1$), they should be readily identifiable at rest-frame IR wavelengths (e.g., \citealt{Yue2013}) with the longer wavebands available with JWST/MIRI.

\begin{table}
\begin{center}
\setlength{\tabcolsep}{1.8mm}
\caption{AGN color-color demarcations}
\begin{tabular}{lccc}

\hline \hline

\multicolumn{1}{c}{$f_{\rm AGN}$}  & 
\multicolumn{1}{c}{F200W-F277W}\tablenote{Each box is demarcated with three or four segments, and each line segment is specified by the F200W-F277W color range over which it holds.}
  & 
\multicolumn{1}{c}{$\alpha$}  & 
\multicolumn{1}{c}{$\beta$} \\
\multicolumn{1}{c}{}  & 
\multicolumn{1}{c}{(AB mag)}  & 
\multicolumn{1}{c}{}  & 
\multicolumn{1}{c}{} \\

\hline
\multicolumn{4}{c}{High Completeness} \\
0.05 &  [0.07, $\infty$] & 	2.5 & 	-0.138 \\	
0.05 & 	[0.07, 0.20] & 	-0.3 & 	0.059 \\
0.05 & 	[0.20, $\infty$] & 	1.2 & 	-0.242 \\

0.10\rule{0pt}{4ex} & 	[0.04, $\infty$] & 	2.5 & 	-0.075 \\
0.10 & 	[0.04, 0.20] & 	-0.4 & 	0.041 \\
0.10 & 	[0.18, $\infty$] & 	1.0 & 	-0.211 \\

0.25\rule{0pt}{4ex} & [0.01, $\infty$] & 	3.0 & 	-0.050 \\
0.25 & 	[0.01, 0.15] & 	-0.3 & 	-0.017 \\
0.25 & 	[0.15, $\infty$] & 	0.8 & 	-0.182 \\

\hline

\multicolumn{4}{c}{\rule{0pt}{4ex}High Purity} \\
0.05 & 	[0.09, $\infty$] & 	2.5 & 	-0.170 \\
0.05 & 	[0.09, 0.25] & 	0.05 & 	0.051 \\
0.05 & 	[0.25, $\infty$] & 	1.6 & 	-0.337 \\

0.10\rule{0pt}{4ex} & 	[0.06, $\infty$] & 	2.2 & 	-0.050 \\
0.10 & 	[0.06, 0.25] & 	-0.2 & 	0.094 \\
0.10 & 	[0.25, $\infty$] & 	1.3 & 	-0.281 \\

0.25\rule{0pt}{4ex} & [0.04, $\infty$] & 	2.2 & 	-0.010 \\
0.25 & 	[0.04, 0.28] & 	-0.2 & 	0.086 \\
0.25 & 	[0.28, $\infty$] & 	1.3 & 	-0.334 \\
\hline
\multicolumn{4}{c}{\rule{0pt}{4ex}Balanced} \\
0.05 & 	[0.09, $\infty$] & 	2.0 & 	-0.090 \\
0.05 & 	[0.09, 0.14] & 	-1.4 & 	0.216 \\
0.05 & 	[0.14, 0.25] & 	0.3 & 	-0.022 \\
0.05 & 	[0.25, $\infty$] & 	1.7 & 	-0.372 \\

0.10\rule{0pt}{4ex} & 	[0.05, $\infty$] & 	2.0 & 	-0.020 \\
0.10 & 	[0.05, 0.13] & 	-1.0 & 	0.130 \\
0.10 & 	[0.13, 0.28] & 	0.4 & 	-0.052 \\
0.10 & 	[0.28, $\infty$] & 	1.5 & 	-0.360 \\

0.25\rule{0pt}{4ex} & [0.025, $\infty$] & 1.9 & 	0.020 \\
0.25 & 	[0.025, 0.14] & -0.8 & 	0.088 \\
0.25 & 	[0.14, 0.28] & 	0.4 & 	-0.081 \\
0.25 & 	[0.28, $\infty$] & 	1.5 & 	-0.389 \\

\hline
\end{tabular}
\end{center}

\normalsize
\vspace{-0.6cm}
\end{table}

\subsection{The Color Box}
\label{sec:colorbox}

Color-color selection boxes are widely used to identify quasar and AGN candidates across a variety of wavelengths, most notably in the observed near-IR and mid-IR bands, such as those using Spitzer and WISE \citep{Lacy:2004,Stern:2005,Mateos:2013}. In the previous section, we noted significant differences between the AGN and pure galaxy populations in the F150W-F200W vs F200W-F277W color-color diagram, which we focus on here to investigate potential color selection boxes that could be used to select BH seed candidates at $z>6$. 

The precise demarcations for AGN selection often depend on the science driver i.e., the requirement of a pure AGN selection for identification vs a more complete AGN selection for population analyses. Hence, we take three approaches here: a selection that provides (1) a 90\% pure AGN sample at the expense of completeness; (2) a 90\% complete AGN sample at the expense of non-AGN interlopers; (3) a balanced purity--completeness AGN selection set by the saddle point between metrics (1) and (2). These metrics, however, require knowledge of the fraction of high redshift galaxies that host an accreting BH ($f_{\rm AGN}$) in order to determine the relative normalizations of the color distributions of the galaxies and AGN. Here we assume three separate AGN fractions, a pessimistic 5\% AGN fraction, a conservative 10\% AGN fraction motivated by the number of Seyfert galaxies observed locally, and a high 25\% fraction, theoretically predicted at similarly high redshifts to those considered here \citep{Volonteri:2017}. The individual galaxies assigned with AGN are randomly drawn from the galaxy population in accordance with the AGN fractions (see Fig.~\ref{fig:colorbox}), and we produce approximate color boxes using linear piece-wise fits to the contours parameterized as ${\rm (F150W-F200W)} = ({\rm F200W-F277W})\alpha_i  + \beta_i$, which we provide in Table 1. We emphasize that based upon the JAGUAR mock, our initial NIRCam flux cuts introduce a negligible fraction ($\sim 0.4$\%) of $z<6$ sources into our overall galaxy selection, but that none of these lower-$z$ interlopers reside within the AGN region explored here.

We emphasize that we do not yet know $f_{\rm AGN}$, a primary goal of future studies will be to determine the number density of these accreting BHs. Hence, we conservatively explore a factor of five in AGN fraction here. We show in Fig.~\ref{fig:colorbox} that our color-box is relatively insensitive to the precise value of $f_{\rm AGN}$. Of course, if fewer than 1-5\% of these galaxies host accreting BHs, then it will prove challenging to find them in the very narrow \jwst\ fields (see further discussion in \S \ref{sec:discussion}).

\section{Discussion}
\label{sec:discussion}

\subsection{{ Comparison with prior work}}

A number of works have investigated the capabilities of \jwst\ to detect growing BHs at high redshift. In terms of the AGN continuum taken alone, our assumed disk model is comparable to that in \citet{Valiante:2018}. Our slope of $\beta = -1.2; f_{\lambda} \propto \lambda^{\beta}$ is similar to their $\beta = -2$, but our treatment of the contribution from the host-galaxy to the observed colors differs substantially. The exact \jwst\ passbands and signatures relied upon in each work vary significantly, and we will not try for a comprehensive review here. However, broadly speaking, there are three kinds of selections that have been discussed in the literature to date, and we compare with each. 

\begin{figure}[t]
\centering
\includegraphics[width=\linewidth]{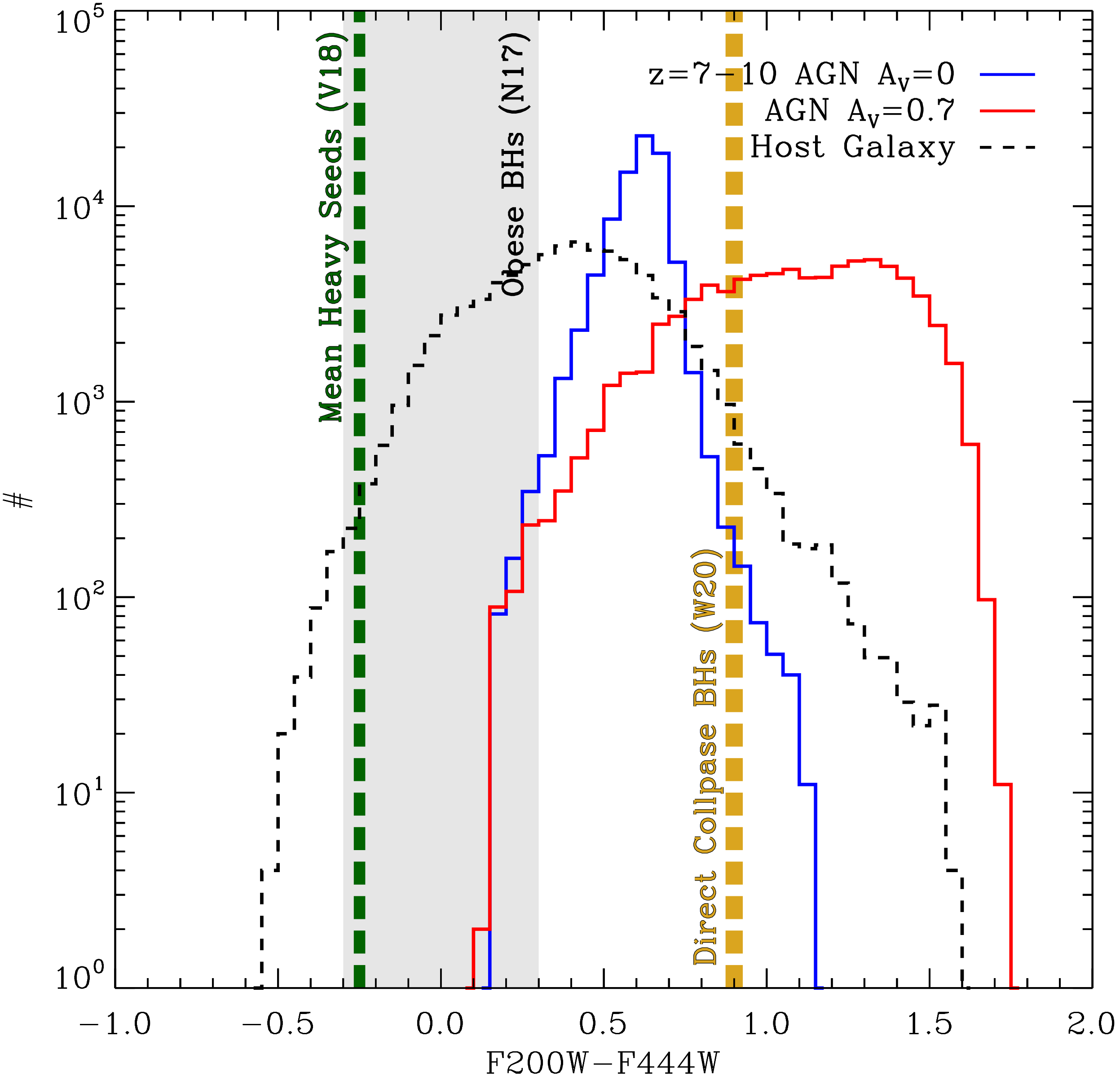}
\caption{Comparison between the F200W--F444W color for JAGUAR z=7--10 AGN+galaxies with no extinction and $A_V = 0.7$ (blue and red histograms, respectively) and pure galaxy emission (black dashed histogram) presented in this work to those found from theoretical BH formation models for obese BHs (Natarajan et al. 2017; N17; gray shaded region), heavy seed mechanisms (e.g., Valiante et al. 2018; N18; green dashed line) and direct collapse BHs (Whalen et al. 2020; W20; yellow dashed line).
\label{fig:comparison}}
\end{figure}

The first is continuum colors in the rest-frame UV. We note that these continuum colors have often been combined with other multi-wavelength/detector indicators (see below), but here we focus exclusively on \jwst/NIRCam. In Fig.~\ref{fig:comparison} we show a comparison in F200W--F444W color-space between the distributions of the galaxies+AGN considered here, and the color range predicted for obese BHs \citep{natarajanetal2017} and the average color of heavy BH seeds \citep{Valiante:2018}. We find that the colors of both heavy seeds and obsese BHs are on the whole bluer (by 0.3--0.8 mags) than those predicted using our $A_{\rm V}=0$ empirical template derived from known NLS1s, and they are both substantially bluer when we include the effect of dust attenuation on our adopted AGN model.  Direct collapse BHs \citep{Whalen:2020} are predicted to have similarly red colors of F200W--F444W$\sim$0.9 to those found in this work. We also note the very blue colors predicted by \citet{Barrow:2018} for direct collapse BHs, albeit at substantially higher redshifts ($z \sim 15$) than we consider here. Furthermore, we find that the F200W--F356W$>$0 cut suggested by \citet{Inayoshi:2022} would similarly select all of the accreting BHs predicted here. However, it must be noted that the distribution of colors occupied by galaxies in JAGUAR is substantially broader and more luminous than those considered in previous studies. Such \jwst\ may be yet further perturbed by the formation of Ultra Compact Dwarfs (e.g., \citealt{Jerabkov:2017}). Based on JAGUAR mock galaxies, we find that the previously proposed color cuts in isolation would likely additionally select a large number of galaxies that are not appreciably growing their BHs at these redshifts.

A second focus of previous studies has been on multi-wavelength searches, where \jwst\ colors are combined with X-ray or far-infrared data \citep{natarajanetal2017,Valiante:2018}. While X-ray data would undoubtedly be valuable, the use of X-rays is at the limit of only the most sensitive field observed with the {\it Chandra} X-ray Observatory. Assuming our fiducial low-mass BH with $M_{\rm BH} = 5 \times 10^6$~\msun at $z=9$ and for a typical $\Gamma=1.7$ X-ray slope and $L_{\rm X} / L_{\rm bol} = 0.1$, we expect $S_X(5-20~{\rm keV,rest}) \gtrsim 10^{-17}$~erg/s/cm$^2$ for $\lambda_{\rm Edd}>0.1$, similar to the X-ray flux limit of the 7~Ms Chandra Deep Field South \citep{Luo:2017}. In practice however, this also relies upon very precise astrometry to exclude extraneous objects detected by \jwst, combined with the smallest available point spread function for the X-ray sources to get above the large instrument background due to the long exposures. Thus, this limits the search area to only the most inner $<1$arcmin$^2$ region of the Chandra Deep Field South. Similar situations arise when considering the requirement of mid to far infrared--X-ray slopes for candidate selections that are beyond the reaches of current infrared survey fields. Without successors to {\it Chandra} or {\it Herschel} on the immediate horizon to provide substantially more sensitive data, it seems prudent to consider \jwst\ selections alone as these data are becoming available now. For example, a further multi-wavelength approach is to combine photometry from both NIRCam and MIRI to access longer wavelengths that are less sensitive to the affects of obscuration but at the expense of sensitivity (e.g., \citealt{Volonteri:2017}) 

Thirdly, a handful of papers consider the power of emission lines in selection, arguing that the extreme equivalent width of H$\alpha$ in the MIRI bands will provide the most unique color signature of accreting BHs in the presence of significant galaxy continuum \citep[e.g.,][]{Inayoshi:2022,Nakajima:2022}. We chose not to include a MIRI component in our selection for now, since the detector is not as sensitive, and there are not (yet) available rest-optical emission line templates available to motivate line strengths.

%%%%%%%%%%%%%%%%%%%%%%%%%%%%%%%%%%
%%%%%%%%%%%%%%%%%%%%%%%%%%%%%%%%%%

\subsection{Expected number densities}

\jwst\ will be sensitive to growing BHs of \mbh$\approx 10^6$~\msun\ and above at very early times ($7 < z < 10$). Current probes of the faint end of the quasar luminosity function reach $M_{\rm UV,1450} \approx -21$~mag, and here we aim to reach as faint as $M_{\rm UV,1450} \approx -16.6$~mag. Given that we have not yet detected AGN of such low luminosity at these redshifts, we address here what reasonable theoretical guesses for the space densities of such BHs might be.

One prediction that we rely on in this paper is the expectation that $\sim 5-25\%$ of galaxies will harbor accreting BHs radiating at $>10\%$ of their Eddington limit \citep{Volonteri:2017}. In that scenario, we expect to detect a large number of low-mass AGN. However, as we go back in time, the number of BHs with \mbh$>10^6$~\msun\ will start to become very sensitive to the seed BH mass. \citet{ricartenatarajan2018} suggest that under heavy seeding models we should expect to detect a few to tens of BHs within $\sim$ hundred square arcmin fields under heavy seed models. In contrast, for light seeds we might not expect any detections, especially at $z>9$.

\subsection{Limitations of this study}

When it comes to modeling the number densities and accretion rates of early growing BHs, the seeding mechanism is not the only major uncertainty \citep[e.g.,][]{ricartenatarajan2018,Inayoshi:2020,Volonteri:2021}. It is also very challenging to implement the dynamics of these low-mass BHs \citep[e.g.,][]{tremmeletal2017,bellovaryetal2019,Piana:2021}, meaning that merger rates, merger timescales, and even the interactions between the low-mass BHs and gas clouds in the proto-galaxies are still quite uncertain. Likewise, without a better understanding of the BH environments, it is likely very challenging to robustly predict accretion rates. In current state-of-the-art hydrodynamical zoom simulations, there are hints that the low-mass BHs simply do not accrete or grow \citep{Latif:2018,bellovaryetal2019}. On the other hand, some papers suggest that the BHs may be able to grow at super-Eddington rates \citep[e.g.,][]{jiangetal2019}. { The launch of \jwst\ raises the exciting prospect that we may start to constrain these uncertainties with measured number densities.}

{ We emphasize that our predicted color boxes are dependent on the assumed galaxy colors from the JAGUAR mock catalog. At present, \jwst\ data are not sufficiently sensitive to validate these color distributions directly, but they will be in the near future. Since the windows to obtain follow-up spectroscopy on the best hi-$z$ BH candidates will be small, it is important to proceed iteratively by obtaining spectra of objects in rare parts of color space immediately, while refining galaxy catalogs in parallel. Furthermore, \jwst\ will be very sensitive for the identification of low-mass central BHs at intermediate redshifts. Hence, these lower redshift BHs will allow future update and validation of our assumed AGN template.}

\section{Summary}

We have presented a simple \jwst/NIRCam color selection designed to tease out low-mass (\mbh$ \approx 10^6$~\msun) BHs embedded in galaxies at $7 < z < 10$. With upcoming deep fields achieving continuum depths of $\sim 30$~AB~mag in NIRCam filters, it is very possible to detect continuum from these accreting BHs, and we argue that it should be possible to distinguish AGN+galaxy light from pure galaxy light, at least in the context of well-motivated mock galaxy catalogs. 

If indeed we cannot observe these growing BHs with radiation, we hope they will reveal themselves via gravitational radiation upon merging \citep[e.g.,][]{bellovaryetal2019,Mangiagli:2022}. In short, new observational constraints, driven by the exquisite sensitivity of \jwst\ are urgently needed to constrain seeding models.

\section*{Acknowledgments}

We thank the anonymous referee for a timely report that allowed us to clarify several important aspects of the manuscript. We are grateful for early discussions about \emph{JWST} with R. Bezanson. J.E.G. and A.D.G acknowledges support from NSF/AAG grant\# 1007094, and J.E.G. also acknowledges support from  NSF/AAG grant \# 1007052.

\bibliographystyle{aasjournal}
\bibliography{imbh.bib}

\end{document}